\documentclass{IAU-TrA}
\usepackage{graphicx}
\usepackage{natbib}

\title[WORKING GROUP ON MOLECULAR DATA] 
{}

\author[DIVISION~XII/COMMISSION~14/WORKING GROUP 
ON MOLECULAR DATA] 
{}

\pubyear{2012}
\volume{Volume XXVIIIA}
\pagerange{001--008}
\jname{Transactions IAU, Volume XXVIIIA}
\editors{Ian Corbett, ed.}
\begin{document}

\maketitle

{\bf

\large
\noindent
DIVISION XII/COMMISSION~14/WORKING GROUP ON                         \\ 
MOLECULAR DATA                                                     \\

\normalsize

\begin{tabbing}
\hspace*{45mm}  \=                                                   \kill
CHAIR           \> Steven R. Federman                                  \\
VICE-CHAIR      \> Peter F. Bernath                                \\
VICE-CHAIR      \> Holger S.P. M\"{u}ller                          \\
\end{tabbing}

\vspace{3mm}

\noindent
TRIENNIAL REPORT 2009-2012
}

\firstsection 

\section{Introduction}

The current report covers the period from the second half of 2003 to the 
first half of 2011, bringing the Working Group's efforts up to date, and is 
divided into three main sections covering rotational, vibrational, and 
electronic spectroscopy. Rather than being exhaustive, space limitations only 
allow us to highlight a representative sample of work on molecular spectra. 
Related research on collisions, reactions on grain surfaces, and astrochemistry 
appear in the report by another Working Group. These also recount recent conferences 
and workshops on molecular astrophysics. 

\section{Rotational Spectra}
\label{Rotational Spectra}

A large number of reviews have appeared dealing with rotational spectra of 
molecules potentially relevant to radio-astronomical observations. Therefore, 
emphasis is placed on investigations dealing with molecular species already 
observed in space. Related molecules are included to a large extent also.
The following groups have been created --  hydride species, anions, molecules 
which may occur in circumstellar envelopes of late type stars, complex 
molecules, weed species, and other molecules -- and are discussed in turn.
\smallskip

Several databases provide rotational spectra of molecular species 
of astrophysical and astrochemical relevance. The two most important sources 
for predictions generated from experimental data by employing appropriate 
Hamiltonian models are the Cologne Database for Molecular Spectroscopy, 
CDMS (http://www.astro.uni-koeln.de/
\newline cdms/) with its catalog 
(http://www.astro.uni-koeln.de/cdms/catalog). An updated description 
appeared in \citet{CDMS_2} and the JPL catalog (http://spec.jpl.nasa.gov/). 
Both also provide primary information, i.e. laboratory data with uncertainties, 
mostly in special archive sections. Additional primary data are available 
in the Toyama Microwave Atlas (http://www.sci.u-toyama.ac.jp/phys/4ken/atlas/).
A useful resource on the detection of certain molecular transitions in space is 
the NIST Recommended Rest Frequencies for Observed Interstellar Molecular 
Microwave Transitions, which has been updated and described by 
\cite{NIST-RRF_2004}.\smallskip

The European FP7 project Virtual Atomic and Molecular Data Centre, VAMDC, 
(http://www.vamdc.org/) aims at combining several spectroscopic, collisional, 
and kinetic databases. The CDMS is the rotational spectroscopy database taking part; 
several infrared databases are also involved. The project has been described 
by \citet{VAMDC_2010}. Other tertiary sources combining data from various databases 
are, e.g., Cassis (http://cassis.cesr.fr/), which provides tools to analyze astronomical 
spectra, or splatalogue (http://www.splatalogue.net/).

\subsection{Hydrides}
\label{hydrides}

Hydrides here are all molecules consisting of one non-metal atom and one or more 
H atoms. They may be neutral or positively charged. Metal hydrides will be dealt 
with in subsection~\ref{CS-mols}. Rotational spectra of some anionic hydride 
molecules have been obtained also, but such molecules have not yet been detected 
in space, see also subsection~\ref{anions}. Hydrides are usually difficult to 
observe from the ground as their fundamental transitions occur in the upper 
millimeter region for heavier species to the terahertz region for lighter ones. 
Some hydrides have recently been detected from the ground, such as $^{13}$CH$^+$ 
with the CSO, or SH$^+$ and OH$^+$ with the APEX telescope. 
The launch of the $Herschel$ satellite in May 2009 has created a wealth of 
opportunities to investigate hydrides, in particular with the high resolution
instrument HIFI. H$_2$O$^+$, H$_2$Cl$^+$, ND, and possibly HCl$^+$ have been 
deteceted, and the fundamental transitions of CH$^+$ and of HF have been 
observed for the first time.\smallskip

The very light hydrides H$_2$D$^+$ and HD$_2 ^+$ have attracted considerable 
attention \citep{H2D+_HD2+_1-1_2005,H2D+_HD2+_1-0_2008,H2D+_THz_2009}. 
Particularly noteworthy is the measurement of the $1_{0,1} - 0_{0,0}$ 
transition of H$_2$D$^+$ more than 60~MHz away from the previously accepted 
value by \citet{H2D+_HD2+_1-0_2008}. A similarly large deviation occurred 
between the initial transition frequency of the $J = 1 - 0$ transition of 
CH$^+$ \citep{CH+_rot_2006} and the one measured by \citet{CH+_rot_2010},
who also recorded the same transition for $^{13}$CH$^+$ and CD$^+$. 
\citet{CH+_analyse_2010} used the latter transition frequencies together with 
data from electronic spectra to derive extensive predictions for rotational 
and rovibrational transitions of several CH$^+$ isotopologues. Isotopic CH 
\citep{CD_13CH_rot_2008}, CH$_2$ \citep{CH2_rot_2005}, CHD \citep{CHD_rot_2011}, 
CH$_2$D$^+$ \citep{CH2D+_rot_2010}, and CH$_3$D \citep{CH3D_rot_2009} have been 
other carbon containing hydrides studied.\smallskip

No accurate rotational transition frequencies are available for CH$_2 ^+$ 
or NH$_2 ^+$. Their fundamental transitions are beyond the region 
which can be studied with HIFI.\smallskip 

The nitrogen hydrides NH \citep{NH_rot_2004}, NH$^+$ \citep{NH+_LMR_2009}, NH$_3$ 
\citep{NH3_rot_2010}, as well as the isotopologues NHD$_2$ \citep{NHD2_rot_2006}, 
$^{15}$NH$_2$D and $^{15}$NHD$_2$ \citep{15NH2D_15NHD2_rot_2008} have been 
investigated recently. Transition frequencies have been determined for the oxygen 
hydrides $^{17}$OH \citep{17OH_LMR_2003}, OH$^+$ \citep{CDMS_2}, H$_2$O and 
H$_2 $$^{18}$O \citep{H2O_H2O-18_LD_2006}, H$_2 $$^{17}$O \citep{H2O-17_LD_2009}, 
D$_2$O \citep{D2O_rot_2007,D2O_LD_2010}, H$_3$O$^+$ \citep{H3O+_2009}, and 
H$_2$DO$^+$ \citep{H2DO+_analyse_2010}.\smallskip

Halogen containing molecules have attracted increased attention in recent years. 
Among the hydride species investigated recently are DF \citep{DF_rot_2006}, 
HF$^+$ \citep{HF+_LMR_2004}, H$_2$F$^+$ \citep{H2F+_rot_2011}, and 
H$_2$Cl$^+$ \citep{H2Cl+_rot_2001}.\smallskip

Other investigations of heavier hydrides include Lamb-dip measurements 
of low-lying rotational states of PH$_3$ \citep{PH3_LD_2006} and a combined 
analysis of spectroscopic data of SH$^+$ \citep{SH+_analysis_2009}.

\subsection{Anions}
\label{anions}

There had been speculation about the possibility of molecular anions in space 
for quite a while. However, accurate transition frequencies were only known 
for OH$^-$ and SH$^-$; for a few other species approximate values were known from 
infrared spectroscopy. However, all these species were deemed to be too 
fragile with respect to photolysis. Moreover, all these species were hydride 
species with fundamental transitions not easily accessible from the ground.
\smallskip

Interestingly, a molecular line survey of the circumstellar envelope of the 
carbon-rich star CW~Leo, also known as IRC~+10216, between 28 and 50~GHz 
revealed several series of unidentified lines \citep{CW-Leo_survey_1995}. 
One of these turned out to be caused by the anion C$_6$H$^-$ \citep{C6H-_rot_2006}. 
This finding sparked a flurry of investigations into laboratory rotational 
spectra of related molecular anions and the search for these species in space. 
C$_2$H$^-$ \citep{C2H-_rot_2007}, as well as C$_4$H$^-$ and 
C$_8$H$^-$ \citep{C4H-_C8H-_rot_2007}, 
were characterized and, except for the smallest member, found in space. 
The isoelectronic molecules CN$^-$ \citep{CN-_rot_2007} and C$_3$N$^-$ 
\citep{C3N-_rot_2008} were found in the lab as well as in the circumstellar 
envelope of CW~Leo. The identification of C$_5$N$^-$ also in that source 
seemed to be certain enough even in the absence of laboratory spectral data. 
Furthermore, submillimeter transitions have been recorded for CN$^-$, C$_2$H$^-$, 
and C$_4$H$^-$ \citep{CN-_C2H-_C4H-_smm_2008} as well as for C$_3$N$^-$ 
\citep{C3N-_smm_2010}. In addition, the laboratory rotational spectrum of NCO$^-$ 
was recently reported  \citep{NCO-_rot_2010}.

\subsection{Circumstellar molecules}
\label{CS-mols}

The chemistry of circumstellar envelopes of late-type stars is special as a 
number of molecules have only been detected or are particularly abundant in 
such sources. Probably the most fascinating object in this regard is CW~Leo.  
Whereas many astronomers have thought this source 
has a rather special chemistry, it turned out that it is a rather common 
source. The reason why molecules are particularly easily detected in this source 
is its proximity. For instance, several metal containing molecules have been detected 
first in the envelope of CW~Leo, but in recent years MgNC, NaCN, and NaCl 
have been detected also in the envelopes of other AGB stars, the latter even 
in those of O-rich AGB stars such as IK~Tau and VY~CMa. Interestingly, recently 
the metal-containing molecules AlO and AlOH as well as PO have been detected 
first in the envelope of VY~CMa. In addition, KCN and FeCN were detected 
toward CW~Leo.\smallskip

For a number of metal-containing molecules rotational spectra have been studied. 
Hydride species include AlH \citep{AlH_D_rot2004}, CrH \citep{CrH_analysis_2006}, 
MnH \citep{MnH_D_rot_2008}, and FeH \citep{FeH_LMR_2006}. Among the nitride and oxide 
species studied are VN and VO \citep{VN_VO_rot2008}, CoO \citep{CoO_rot_2005}, ZnO 
\citep{ZnO_rot_2009}, and even TiO$_2$ \citep{TiO2_rot_2011}. Several halogenides 
have been investigated, such as NaCl and KCl \citep{KCl_rot_2004}, TiF 
\citep{TiF_rot_2003}, TiCl \citep{TiCl_rot_2001}, VCl  \citep{VCl_rot_2009}, 
CoF \citep{CoF_rot_2007}, CoCl \citep{CoCl_rot_2004}, ZnF \citep{ZnF_rot_2006}, 
and ZnCl \citep{ZnCl_rot_2007}. Also recorded were rotational spectra of 
CoCN and NiCN \citep{CoCN_rot_2004,NiCN_rot_2003}.\smallskip

The study of the rotational spectrum of C$_2$P \citep{CCP_rot_2009} laid the ground 
work for its detection in the circumstellar envelope of CW~Leo. This detection, in turn, 
suggests that molecules such as C$_2$F, C$_2$Cl, C$_3$F, and C$_3$Cl may be 
detectable as well. Rotational spectra have been detected for all of these 
species except for the first one \citep{CCCl_rot_2003,C3F_rot_2009,C3Cl_rot_2009}.
\smallskip 

Other molecules investigated, including predominantly or potentially circumstellar ones, 
are C$_2$H and C$_6$H in excited vibrational states \citep{{C2H_vib_2007},C6H_vib_2010} 
and CH$_3$CP \citep{CH3CP_rot_2003}.

\subsection{Complex molecules}
\label{complex}

Some of the smaller complex molecules with many emission or absorption lines 
observable by radio-astronomical means are presented in subsection~\ref{weeds}.\smallskip

Among the larger saturated or almost saturated molecules, propenal, propanal, 
and acetamide have been detected with the GBT in the microwave region. 
Interestingly, propylene has been detected with the IRAM 30\,m telescope 
toward TMC-1. Three complex molecules have been detected with the same 
instrument in the course of a molecular line survey of Sagittarius~B2(N) 
at 3\,mm. These are aminoacetonitrile, $n$-propyl cyanide, and ethyl formate. 
The detection articles also feature critical evaluations of the spectroscopic 
parameters of the former two molecules \citep{det_AAN_2008,det_n-PrCN_EtFo_2009}. 
The rotational spectrum of ethyl formate has also been revisited 
\citep{EtFo_rot_2009}. Other studies include $cyclo$-propyl cyanide 
\citep{c-PrCN_rot_2008} and $iso$-propyl cyanide \citep{i-PrCN_rot_2011}, 
two conformers of ethylene glycol \citep{aGg'_2003,gGg'_rot_2004}, microwave 
spectra of several conformers of 1,2- and 1,3-propanediol 
\citep{1-2-Pr2OH_2009,1-3-Pr2OH_2009}, propane \citep{C3H8_rot_2006}, 
millimeter wave spectra of the amino acids glycine and alanine 
\citep{glycine_mmw_2005,alanine_mmw_2008}, acetamide \citep{acetamide_rot_2004}, 
methylamine \citep{CH3NH2_2007}, acetic acid \citep{HAc_2008}, $n$-propanol 
\citep{n-PrOH_rot_2010}, and diethyl ether \citep{DEE_2009}.
Other investigations may turn out to also be relevant once telescope arrays 
such as EVLA, NOEMA, or ALMA conduct large scale line surveys or dedicated 
seaches for particular molecules.

\subsection{Weed species}
\label{weeds}

The term ``weed species'' has been coined for molecules that have very many 
emission or absorption lines in various sources, but considered mostly for 
star-forming regions. It should be emphasized that the plethora of lines is not 
only a nuisance, but there is also considerable information about, for example, 
temperature or density in these lines. The molecule with particularly many 
rather strong lines is methanol. It has been extensively studied up to 
the second torsional state \citep{MeOH_rot_2008}; even higher states have been 
studied to some extent \citep{MeOH_vt3_2009}, but these are difficult to model 
at present. CH$_3$OH has also been proposed as a particularly well suited 
molecule to investigate the possibility of temporal or spatial variations 
of fundamental constants \citep{me-mp_2011a,me-mp_2011b}. One requirement is 
the knowledge of very highly accurate transition frequencies. Some of these 
have been summarized as well as newly reported ones by \citet{MeOH_rot_2004}. 
Other methanol isotopologues studied recently include CH$_2$DOH 
\citep{CH2DOH_rot_2002,CH2DOH_FIR_2009}, CH$_3$OD \citep{MeOD_rot_2003}, and 
CH$_3$$^{18}$OH \citep{18O_MeOH_analysis_2007}.\smallskip

Other weed molecules, for which several isotopic species have been investigated, 
are methyl cyanide \citep{MeCN_rot_2009}, ethyl cyanide 
\citep{EtCN_rot_2009,13C-EtCN_rot_2007,D_N-EtCN_rot_2009}, vinyl cyanide 
\citep{VyCN_rot_2009}, methyl formate 
\citep{MeFo_rot_2009,13C2_MeFo_rot_2009,13C1_MeFo_rot_2010,D-Me-MeFo_rot_2009,
D-Ac-MeFo_rot_2010}, formamide \citep{formamide_2009}, and 
acetone \citep{acetone_vib_2008,13C-acetone_2006}. 
Moreover, dimethyl ether \citep{DME_rot_2009}, its isomer ethanol \citep{EtOH_rot_2008}, 
and the only inorganic weed species SO$_2$ \citep{SO2_rot_2005} have been studied.
\smallskip 

One should keep in mind that weed species are not only seen in the ground 
vibrational state, but often in several excited states, and this is true for minor 
isotopic species as well. While deriving an appropriate Hamiltonian model is usually 
the most compact form to represent the rotational spectrum of a molecule in 
a given vibrational state, this task can be formidable and worse in cases 
of strong vibration-rotation interaction that may take much more than months 
to be tackled. As an alternative, the De Lucia group proposed recording spectra 
at different temperatures. While this procedure yields huge amounts of data, and 
extrapolation in frequency or to higher temperatures is impossible and 
extrapolation to lower temperatures only to some extent, it seems to be still 
a pragmatic approach for some molecules. Several reports have been published 
on ethyl cyanide \citep{EtCN-T_rot_2010} and one on vinyl cyanide 
\citep{VyCN-T_rot_2011}.

\subsection{Other molecules}
\label{other}

Other molecules, for which rotational spectra have been (re-) investigated, include 
cations such as HCNH$^+$ 
and CH$_3$CNH$^+$ \citep{HCNH+_CH3CNH+_rot_2006}, CS$^+$ 
\citep{CS+_rot_2008}, CF$^+$ \citep{CF+_rot_2010}, $^{15}$N isotopologues of 
N$_2$H$^+$ \citep{15N-N2H+_rot_2009}, HCS$^+$ \citep{HCS+_rot_2003}, and HCO$^+$ 
\citep{HCO+_rot_2007}.\smallskip 

Other short-lived molecules include DNC \citep{DNC_rot_2006}, C$_3$H and $^{13}$C 
isotopologues of C$_n$H with $n = 3, 5-7$ \citep{C3H_rot_2009,13C-C3H_etc_2005}, 
HOCN, HONC, and HSCN \citep{HOCN_rot_2009,HONC_rot_2009,HSCN_rot_2009} as well 
as SiN and PN \citep{SiN_rot_2006,PN_rot_2006}.\smallskip

In addition, several stable or fairly stable molecules have been studied, 
including isotopologues of CO in their ground vibrational states \citep{13C17O_13C18O_2003} 
and the main isotopologue in excited vibrational states \citep{CO_vib_2009}, various 
isotopologues in the ground and excited states of CS and SiS \citep{CDMS_2,SiS_rot}, 
isotopologues of HCN in their ground vibrational states \citep{H13CN_LD_2005} 
and the main isotopologue in excited vibrational states \citep{HCN_vib_2003}, 
H$_2$CO \citep{H2CO_rot_2003}, the isoelectronic H$_2$CNH \citep{H2CNH_rot_2010}, 
H$_2$CS \citep{H2CS_rot_2008}, isotopologues of HCOOH \citep{HCOOH_isos_2008}, 
and CH$_3$C$_2$H and CH$_3$C$_4$H \citep{MeC2_4H_LD_2008}.

\section{Vibrational Spectra}
\label{Vibrational Spectra}
Here we describe vibration-rotation spectra of gaseous molecules of astronomical or 
potential astronomical interest. It is based in part on a review article entitled 
``Molecular astronomy of cool stars and sub-stellar objects'' aimed at physical 
chemists and laboratory spectroscopists \citep{Bernath2009}.\smallskip

In addition to the references to particular molecules given below there are a number of 
spectral database compilations that are useful. Perhaps the most helpful is the HITRAN 
database that contains vibration-rotation line parameters for a large number of species 
such as H$_2$O, CO$_2$, CO, etc. found in the Earth's atmosphere \citep{HITRAN2008}. 
HITRAN is widely used for astronomical applications although it is not always suitable 
because of missing lines and bands, particularly in the near infrared region. For example, 
``cool'' astronomical objects such as brown dwarfs can have surface temperatures in excess 
of 1000 K, and HITRAN is designed for temperatures near 300 K. In this regard, there is a 
HITEMP database \citep{HITEMP2010} for H$_2$O, CO$_2$, CO, NO, and OH that is more 
suitable for high temperature sources such as stellar atmospheres.\smallskip

For larger molecules, individual vibration-rotation lines are no longer clearly resolved 
and it becomes necessary to replace line-by-line calculations by absorption cross 
sections. The main drawback to using cross sections is that a considerable number of 
laboratory measurements are needed to match the temperature and pressure conditions of 
the objects under observation. HITRAN also includes a number of high resolution infrared 
absorption cross sections for organic molecules such as ethane and acetone, but the 
broadening gas in HITRAN is air rather than H$_2$, N$_2$, or CO$_2$. While the GEISA 
database has significant overlap with HITRAN, it contains additional molecules of 
interest for studies of planetary atmospheres \citep{jac11}. 
A very useful set of infrared absorption cross sections for several hundred molecules 
have been measured at the Pacific Northwest 
National Laboratory (PNNL) for the 600-6500 cm$^{-1}$ (1.54--16.7 $\mu$m) range 
\citep{PNNL2004}. The PNNL IR database, however, may not be completely suitable for 
astronomical applications, for example for planetary atmospheres. All PNNL spectra are 
recorded at relatively low resolution (0.112 cm$^{-1}$) as mixtures with pure nitrogen gas 
at pressures of 760 Torr and temperatures of 278, 293, or 323 K. Nevertheless, in the 
absence of spectra recorded under more appropriate experimental conditions, they can be 
very useful.
\smallskip

Other interesting general sources for infrared data are various high resolution spectral 
atlases of the Sun \citep{Solar2003, ACE2010} and 
sunspots \citep{Sunspot2001, Sunspot2002} 
because they include molecular (and atomic) line assignments. The web site spectrafactory 
\citep{cami2010} is also useful for calculating infrared spectra of astronomical interest, 
although not all of the input line parameters are the most recent or recommended ones.

\subsection{Diatomic Molecules}
\label{Diatomic Molecules}

Diatomic molecules, particularly diatomic hydrides of the more abundant elements, are 
often observed through their infrared spectra. Molecular hydrogen itself is difficult to 
observe in the infrared because all transitions are electric-dipole forbidden; however, 
HD has a small dipole moment and very recently the 2--0 band was measured near 
1.4 $\mu$m by cavity ring down spectroscopy \citep{HD2011}. In the case of OH, NH, 
and CH, the new infrared solar atlas \citep{ACE2010} measured by the Atmospheric 
Chemistry Experiment (ACE) infrared Fourier transform 
spectrometer from orbit \citep{ACE2005} has led to an 
improvement in the spectroscopy. The infrared Fraunhofer lines were combined with 
laboratory measurements to extend the measured line positions to higher vibrational and 
rotational quantum numbers for OH \citep{OH2009}, NH \citep{NH2010}, and CH 
\citep{CH2010}.\smallskip

Laboratory vibration-rotation spectra for a considerable number of metal hydrides are 
known with recent measurements of, for example, BeH \citep{BeH2003}, MgH \citep{MgH2004}, 
and CaH \citep{CaH2004}; even the metal dihydrides BeH$_2$ \citep{BeH2-2003} and 
MgH$_2$ \citep{MgH2-2003} have been detected in the laboratory. Although metal hydrides 
such as MgH and CaH are detected in stellar atmospheres by their electronic transitions, 
there have been no infrared astronomical observations. Other diatomics such as CO, SO, 
SiO, HF, HCl, SH, CS, and SiS are commonly seen by infrared observations of cool stellar 
atmospheres \citep{Bernath2009}; their spectroscopy however has not been improved much 
in recent years. An exception to this is some work on the line intensities of SiS 
\citep{cami2009} and HCl \citep{gang2011}.

\subsection{Small Polyatomics}
\label{Small Polyatomics}

The spectra of ``cold'' water, ammonia, and methane as given in the HITRAN database are 
generally satisfactory for astronomical purposes, 
except for overtone and combination bands of NH$_3$ and CH$_4$ in the near 
infrared and visible regions. For spectra of hot samples as 
in brown dwarfs and exoplanets, the situation is much less sanguine, except in the case of 
water for which there has been extensive experimental \citep{steam2008} and theoretical 
work \citep{BT2}. Much work is continuing on highly-excited water levels near dissociation 
and on computing line intensities ab initio, 
but the BT2 line list \citep{BT2} is generally suitable for computing water opacities. 
For hot ammonia there has been rapid recent progress with new laboratory spectra 
recorded \citep{York2011} and with at least two groups providing 
rather good calculated spectra \citep{Ames2011,UCL2011}. Ammonia is seen in brown dwarfs 
and is thought to be the key molecule in defining a new class of ultracool Y-type dwarfs 
\citep{Y-dwarfs2011}. Methane is the laggard because of the difficulty of the problem, and 
the existing experimental data \citep{Nassar2003} 
and calculations are not very satisfactory 
for simulating brown dwarf and exoplanet spectra. For CO$_2$, HITRAN for cold molecules 
and HITEMP or CDSD-4000 databases \citep{cdsd} for hot molecules are recommended. 

\subsection{Large Molecules}
\label{Large Molecules}

The discovery of C$_{60}$ and C$_{70}$ in the young planetary nebula Tc1 was reported 
by \citet{cami-C60-2010} using the Spitzer Space Telescope. The discovery was rapidly 
confirmed with detections in other planetary nebulae; some of these sources also have 
spectral features generally attributed to polycyclic aromatic hydrocarbons (PAHs). 
Infrared emission spectra of gaseous C$_{60}$ and C$_{70}$ were recorded as a function of 
temperature by \citet{Nemes1994} and more recently the integrated molar absorptivity of 
the infrared bands of the solid has been measured as a function of temperature 
\citep{solid2011}. There have also been extensive laboratory measurements and quantum 
chemical calculations of PAH spectra, some of which is summarized in the NASA Ames 
database \citep{ames}. Traditionally infrared spectra of neutral and ionized PAHs have 
been recorded by matrix isolation spectroscopy, but now gas-phase measurements are 
possible \citep{duncan2009, galue2011}. Computational studies of PAHs and related species 
have also advanced, and the quality and quantity of calculations available is remarkable 
\citep{ames2,ames}.

\section{Electronic Spectra}
\label{Electronic Spectra}

In this section we describe recent work on electronic spectra that includes line 
identification, energy levels, as well as data needed for photochemical models. The 
data come in a variety of forms, such as absorption cross sections (or equivalently 
oscillator strengths), predissociation widths, and analyses of line anomalies 
resulting from perturbations between energy levels. Here we can only present a 
representative sampling of work on electronic spectra. The number of experiments on 
electron excitation/scattering is great, and we provide a few illustrative examples 
of research in this area. This section is divided into three topics: interstellar 
matter, which includes diffuse molecular clouds and disks around newly formed 
stars as well as comets whose chemistry is similar, metal hydrides and oxides in 
the spectra of late-type stars, and the atmospheres of planets and their satellites. 
Recent attempts to identify the diffuse interstellar bands is not included, but 
instead we refer the reader to a review on laboratory astrophysics \citep{sav11} that 
includes a discussion of this topic.
It is worthwhile noting, however, that the study of the diffuse interstellar bands has 
led to a renaissance in the gas phase spectroscopy of complex carbon molecules 
that begun in 2003. A typical example is the case of PAH molecules. Traditionally 
UV-Visible spectra of neutral and ionized PAHs could only be recorded by matrix 
isolation spectroscopy, but now gas-phase measurements are possible (see reviews 
in \cite{joblin11}).

\subsection{Interstellar matter}
\label{Interstellar matter}

There is continued interest in spectroscopic studies of CO and much of the recent 
efforts are needed for improved photochemical modeling. A comprehensive analysis of 
Rydberg states has been published \citep{eid04a}, as has a compilation of 
triplet and singlet transitions \citep{eid03} that included oscillator strengths. 
Improved wavelengths have appeared 
for the $A$ $-$ $X$ system of bands in C$^{17}$O and C$^{18}$O 
\citep{ste03,dup06,dup07}, for the $E$ $-$ $X$ (0,0) band in CO, $^{13}$CO, and 
$^{13}$C$^{18}$O \citep{cac04}, and for triplet-singlet bands involving the $e$, $d$, 
and $a^{^\prime}$ states in several isotopologues \citep{dup07,yan08,dic10}. 
Several quantum mechanical calculations of potential energy curves 
\citep{cha06,vaz09,lef10} have appeared. Empirical determinations of oscillator 
strengths for Rydberg transitions have been published, including one based on 
interstellar spectra acquired with the {\it Far Ultraviolet Spectroscopic 
Explorer} \citep{she03}, two using synchrotron radiation \citep{eid04b,eid06}, 
and one employing electron scattering techniques \citep{kaw08}. \citet{gil07} 
have obtained the lifetime for the $a$ $^3\Pi$ $v=0$ level. In their study, 
\citet{eid06} have also reported predissociation rates for $B$ and $W$ states; 
the large rates found for the $B$ $^1\Sigma^+$ $v=6$ level result from 
interactions with the $D^{\prime}$ state. A number of other experimental and 
theoretical efforts describing the $B$ $-$ $D^{\prime}$ interaction have been 
published \citep{and04,gro04,bak05a,bit07}. The triplet $k$ and $c$ states 
have been the focus of studies \citep{bak05b,bakl05a,bakl05b,bakl05c} on 
perturbations caused with other states, while \citet{ben07} have described their 
experiment on perturbations between the $a$ and $d$ levels. Finally, a review 
\citep{lef07} discussing perturbations in the isoelectronic molecules, CO and 
N$_2$, has also appeared. More spectroscopic work on N$_2$ is provided below.
\smallskip

New spectroscopic studies on CH, CH$^+$, NH have appeared since 2003. Extending the 
study of \citet{wat01} on Rydberg transitions in CH, \citet{she07} have inferred 
oscillator strengths and predissociation rates for the 3$d$ $-$ $X$, 4$d$ - $X$, 
$F$ $-$ $X$, and $D$ $-$ $X$ bands seen in interstellar spectra acquired with the 
{\it Hubble Space Telescope}. Theoretical efforts on CH include structure 
calculations of the 3$d$ complex \citep{vaz07} and oscillator strengths for 
Rydberg transitions \citep{lav09}; the latter study also determined photoionization 
cross sections. A new analysis of the $A$ $^1\Pi$ $-$ $X$ $^1\Sigma^+$ system in 
CH$^+$ has been completed \citep{hak06}, and \citet{wes09} have obtained oscillator 
strengths for the (1,0), (2,0), (3,0), and (4,0) bands of this system from 
ground-based interstellar spectra. CH$^+$ photodissociation cross sections have 
been a focus of recent theoretical work \citep{bar04,bou05}. Revised term values 
have been determined for the $X$ $^3\Sigma^-$ and $A$ $^3\Pi$ states of NH by 
\citet{NH2010}, who analyzed spectra from the ACE and ATMOS instruments.\smallskip

Other molecules of interest to interstellar and cometary studies, such as 
C$_2$, CN, and C$_3$, have been studied recently. Theoretical efforts have 
computed oscillator strengths and radiative lifetimes for singlet (Phillips) and 
triplet (Swan, Ballik-Ramsay, and $d$ $-$ $c$) systems in C$_2$ \citep{kok07,sch07}; 
of particular note is that a self-consistent set of oscillator strengths for the 
Phillips ($A$ $-$ $X$) system of bands is emerging. Experimental studies involving 
triplet states have been performed with a variety of techniques 
\citep{joe07,tan07,nak09,bor10,bor11}, many times focusing on the presence of 
perturbations. \citet{tof04} have calculated near threshold photoionization cross 
sections for C$_2$. Spectroscopic studies of the violet and red band 
systems for isotopologues of CN \citep{ram06,ram10b,ram10c} have improved the 
precision of the molecular constants for this molecule. This work will also be of 
interest to stellar astronomy. \citet{shi10} have obtained spectroscopic 
parameters in the isotopologues of CN through {\it ab initio} calculations. Moreover, 
experimental work on bands in the $A$ $^1\Pi_u$ $-$ $X$ $^1\Sigma^+_g$ system of 
C$_3$ has appeared \citep{mcc03,tan05,zha05,che10,che11}. Perturbations have been 
seen in many of these bands. \citet{zha05} have measured lifetimes as well.
\smallskip

Spectroscopic studies on H$_2$O and HCl have also been conducted. \citet{fil04} 
used synchrotron radiation to examine the intense Rydberg $nd$ series in H$_2$O for 
energies up to 12 eV. Another study with synchrotron radiation found several new 
vibrational progressions around 10 eV and also reported abolute cross sections 
\citep{mot05}. Additional experimental efforts obtained absolute absorption cross 
sections with a synchrotron source \citep{che04} and oscillator strengths via 
electron impact-excitation \citep{tho07}; \citet{che04} compared their results 
with additional theoretical computations. \citet{bor06b} extended his earlier 
calculations \citep{bor06a} in a study of oscillator strengths for the {\it \~{A}} 
$^1B_1$ $-$ {\it \~{X}} $^1A_1$ transition in H$_2$O. Another electron scattering 
experiment \citep{li06} derived oscillator strengths for valence-shell 
excitations in HCl.

\subsection{Late-type stars}
\label{Late-type stars}

The transition from metal oxides to metal hydrides is a signature of the latest 
stellar spectral types.  During the reporting 
period papers on line lists, line strengths, 
and opacties have appeared. For metal hydrides, these studies include high-resolution 
spectra acquired with a Fourier transform spectrometer of bands in the 
$A$ $^2\Pi$ $-$ $X$ $^2\Sigma^+$ and $B$ $^2\Sigma^+$ $-$ $X$ $^2\Sigma^+$ systems of 
$^{24}$MgH \citep{sha07} and of the $E$ $^2\Pi$ $-$ $X$ $^2\Sigma^+$ transition in 
CaH and CaD \citep{ram11b}. The work on MgH provided the data on the highest lying 
vibrational level in the ground electronic state. A deperturbation analysis of the 
MgH spectra was recently completed \citep{sha11}. Using laser-induced fluorescence, 
\citet{cho06} examined the (1,0) band of the $A$ $^6\Sigma^+$ $-$ $X$ $^6\Sigma^+$ 
transition in CrH. New theoretical calculations have led to improved line lists and 
opacities for transitions in TiH \citep{bur05} and the $F$ $^4\Delta_i$ $-$ 
$X$ $^4\Delta_i$ transitions in FeH \citep{dul03}. Magnetic properties of FeH 
involving the $F$ and $X$ states, including a study of the Zeeman effect, have been 
determined \citep{har08a,har08}.

\subsection{Planetary atmospheres}
\label{Planetary atmospheres}

Nitrogen and sulfur dioxide have received a considerable amount of 
attention recently. Spectroscopic work on N$_2$ 
\citep{spr03,spr05b,lew08a,vie08} has focused on extreme UV transitions, 
where perturbations and predissociation 
play a significant role; the study by \citet{spr03} considered the 
$^{15}$N$_2$ and $^{14}$N$^{15}$N isotopologues. 
A theoretical calculation \citep{lew08b} has studied the perturbations 
involving these high-lying Rydberg states. Oscillator strengths 
and line widths have been obtained from a number of experimental 
\citep{sta05,sta08,hea09,hub09} and theoretical \citep{jun03,hav05,lav11} 
efforts. Lifetimes of Rydberg states have been determined through a 
combination of experiments and calculations 
\citep{spr04,lew05a,lew05b,spr05a,spr06}. As for SO$_2$, the focus has 
been on absorption cross sections at UV wavelengths 
\citep{ruf03,ruf09,dan08,her09,bla11}, where \citet{dan08} studied 
isotolopogues for sulfur. Furthermore, absorption cross sections for 
ammonia and its isotopologues \citep{che06,wu07} and carbon dioxide 
\citep{sta07} have been determined experimentally.

\vspace{3mm}
 
{\hfill Steven R. Federman}

{\hfill {\it chair of Working Group}}


\begin{thebibliography}{}

\bibitem[Allen et al.(2004)]{HF+_LMR_2004} 
Allen, M.D., Evenson, K.M., \& Brown, J.M. 
2004, J. Mol. Spectrosc., 227, 13
\bibitem[Amano \& Hirao(2005)]{H2D+_HD2+_1-1_2005} 
Amano, T., \& Hirao, T. 
2005, J. Mol. Spectrosc., 233, 7 
\bibitem[Amano et al.(2006)]{HCNH+_CH3CNH+_rot_2006} 
Amano, T., Hashimoto, K., \& Hirao, T. 
2006, J. Mol. Struct., 795, 190
\bibitem[Amano(2008)]{CN-_C2H-_C4H-_smm_2008} 
Amano, T. 
2008, J. Chem. Phys., 129, 244305
\bibitem[Amano(2010a)]{CH+_rot_2010} 
Amano, T. 
2010a, ApJ, 716, L1 
\bibitem[Amano(2010b)]{CH2D+_rot_2010} 
Amano, T. 
2010b, A\&A, 516, L4
\bibitem[Amano(2010c)]{C3N-_smm_2010} 
Amano, T. 
2010c, J. Mol. Spectrosc., 259, 16
\bibitem[Andric et al.(2004)]{and04}
Andric, L., Bouakline, F., Grozdanov, T.P., \& 
McCarroll, R. 2004, 
A\&A, 421, 381
\bibitem[Araki et al.(2001)]{H2Cl+_rot_2001} 
Araki, M., Furuya, T., \& Saito, S. 
2001, J. Mol. Spectrosc., 210, 132
\bibitem[Asvany et al.(2008)]{H2D+_HD2+_1-0_2008} 
Asvany, O., Ricken, O., M{\"u}ller, H.S.P., Wiedner, M.C., 
Giesen, T.F., \& Schlemmer, S. 
2008, Phys. Rev. Lett., 100, 233004
\bibitem[Bailleux et al.(2008)]{CS+_rot_2008} 
Bailleux, S., Walters, A., Grigorova, E., \& Margul{\`e}s, L. 
2008, ApJ, 679, 920
\bibitem[Baker(2005a)]{bak05a}
Baker, J., 2005a, 
Chem. Phys. Lett., 408, 312
\bibitem[Baker(2005b)]{bak05b}
Baker, J., 2005b, 
J. Mol. Spectrosc., 234, 75
\bibitem[Baker \& Launay(2005a)]{bakl05a}
Baker, J., \& Launay, F. 2005a,
Chem. Phys. Lett., 404, 49
\bibitem[Baker \& Launay(2005b)]{bakl05b}
Baker, J., \& Launay, F. 2005b,
Chem. Phys. Lett., 415, 296
\bibitem[Baker \& Launay(2005c)]{bakl05c}
Baker, J., \& Launay, F. 2005c,
J. Chem. Phys., 123, 234302
\bibitem[Barber et al.(2006)]{BT2}
Barber, R.J., Tennyson, J., Harris, G.J., \& Tolchenov, R.N. 2006, 
MNRAS, 368, 1087
\bibitem[Barinovs \& van Hemert(2004)]{bar04}
Barinovs, G., \& van Hemert, M.C. 2004, 
Chem. Phys. Lett., 399, 406
\bibitem[Bauschlicher et al.(2010)]{ames}
Bauschlicher, C.W., et al. 2010, 
ApJS, 189, 341. See http://www.astrochem.org/pahdb/
\bibitem[Belloche et al.(2008)]{det_AAN_2008} 
Belloche, A., Menten, K.M., Comito, C., M{\"u}ller, H.S.P., 
Schilke, P., Ott, J., Thorwirth, S., \& Hieret, C. 
2008, A\&A, 482, 179
\bibitem[Belloche et al.(2009)]{det_n-PrCN_EtFo_2009} 
Belloche, A., Garrod, R.T., M{\"u}ller, H.S.P., Menten, 
K.M., Comito, C., \& Schilke, P. 
2009, A\&A, 499, 215
\bibitem[Ben et al.(2007)]{ben07}
Ben, J., Li, L., Zheng, L., Chen., Y., \& Yang, X. 2007, 
Chem. Phys., 335, 109
\bibitem[Bernath et al.(2005)]{ACE2005}
Bernath, P.F., et al. 2005, 
Geophys. Res. Lett., 32, L15S01. See http://www.ace.uwaterloo.ca/
\bibitem[Bernath(2009)]{Bernath2009}
Bernath, P.F. 2009, 
Int. Rev. Phys. Chem., 28, 681
\bibitem[Bernath \& Colin(2009)]{OH2009}
Bernath, P.F., \& Colin, R. 2009, 
J. Mol. Spectrosc., 257, 20
\bibitem[Bitencourt et al.(2007)]{bit07}
Bitencourt, A.C.P., Prudente, F.V., \& Vianna, J.D.M. 2007, 
J. Phys. B., 40, 2075
\bibitem[Bizzocchi et al.(2003)]{CH3CP_rot_2003} 
Bizzocchi, L., Cludi, L., \& Degli Esposti, C. 
2003, J. Mol. Spectrosc., 218, 53
\bibitem[Bizzocchi et al.(2006)]{SiN_rot_2006} 
Bizzocchi, L., Degli Esposti, C., \& Dore, L. 
2006, A\&A, 455, 1161
\bibitem[Bizzocchi et al.(2008)]{c-PrCN_rot_2008} 
Bizzocchi, L., Degli Esposti, C., Dore, L., \& Kisiel, Z. 
2008, J. Mol. Spectrosc., 251, 138
\bibitem[Blackie et al.(2011)]{bla11}
Blackie, D., Blackwell-Whitehead, R., Stark, G., Pickering, 
J.C., Smith, P.L., Rufus, J., \& Thorne, A.P. 2011, 
JGRE, 116, 03006
\bibitem[Borges(2006a)]{bor06a}
Borges, I. 2006a, 
J. Phys. B, 39, 641
\bibitem[Borges(2006b)]{bor06b}
Borges, I. 2006b, 
Chem. Phys., 328, 284
\bibitem[Bornhauser et al.(2010)]{bor10}
Bornhauser, P., Knopp, G., Gerber, T., Radi, P.P. 2010, 
J. Mol. Spectrosc., 262, 69
\bibitem[Bornhauser et al.(2011)]{bor11}
Bornhauser, P., Sych, Y., Knopp, G., Gerber, T., \& Radi, 
P.P. 2011, 
J. Chem. Phys., 134, 044302
\bibitem[Bouakline et al.(2005)]{bou05}
Bouakline, F., Grozdanov, T.P., Andric, L., \& McCarroll, R. 
2005, J. Chem. Phys., 122, 044108
\bibitem[Brauer et al.(2009)]{EtCN_rot_2009} 
Brauer, C.S., Pearson, J.C., Drouin, B.J., \& Yu, S. 
2009, ApJS, 184, 133
\bibitem[Brown et al.(2006)]{FeH_LMR_2006} 
Brown, J.M., K{\"o}rsgen, H., Beaton, S.P., \& Evenson, K.M. 
2006, J. Chem. Phys., 124, 234309
\bibitem[Brown \& M{\"u}ller(2009)]{SH+_analysis_2009} 
Brown, J.M., \& M{\"u}ller, H.S.P. 
2009, J. Mol. Spectrosc., 255, 68
\bibitem[Br{\"u}nken et al.(2003)]{H2CO_rot_2003} 
Br{\"u}nken, S., M{\"u}ller, H.S.P., Lewen, F., \& Winnewisser, G. 
2003, Phys. Chem. Chem. Phys., 5, 1515
\bibitem[Br{\"u}nken et al.(2005)]{CH2_rot_2005} 
Br{\"u}nken, S., M{\"u}ller, H.S.P., Lewen, F., \& Giesen, T.F. 
2005, J. Chem. Phys., 123, 164315
\bibitem[Br{\"u}nken et al.(2006)]{DNC_rot_2006} 
Br{\"u}nken, S., M{\"u}ller, H.S.P., Thorwirth, S., Lewen, F., 
\& Winnewisser, G. 
2006, J. Mol. Struct., 780, 3
\bibitem[Br{\"u}nken et al.(2007a)]{D2O_rot_2007} 
Br{\"u}nken, S., M{\"u}ller, H.S.P., Endres, C., Lewen, F., 
Giesen, T., Drouin, B., Pearson, J.C., M{\"a}der, H. 
2007a, Phys. Chem. Chem. Phys., 9, 2103 
\bibitem[Br{\"u}nken et al.(2007b)]{C2H-_rot_2007} 
Br{\"u}nken, S., Gottlieb, C.A., Gupta, H., McCarthy, M.C., 
\& Thaddeus, P. 
2007b, A\&A, 464, L33
\bibitem[Br{\"u}nken et al.(2009a)]{HOCN_rot_2009} 
Br{\"u}nken, S., Gottlieb, C.A., McCarthy, M.C., \& Thaddeus, P. 
2009a, ApJ, 697, 880
\bibitem[Br{\"u}nken et al.(2009b)]{HSCN_rot_2009} 
Br{\"u}nken, S., Yu, Z., Gottlieb, C.A., McCarthy, M.C., 
\& Thaddeus, P. 
2009b, ApJ, 706, 1588
\bibitem[Burrows et al.(2005)]{bur05}
Burrows, A., Dulick, M., Bauschlicher, C.W., Bernath, P.F., 
Ram, R.S., Sharp, C.M., \& Milsom, J.A. 2005,
ApJ, 624, 988
\bibitem[Cacciani \& Ubachs(2004)]{cac04}
Cacciani, P., \& Ubachs, W. 2004, 
J. Mol. Spectrosc., 225, 62
\bibitem[Cami et al.(2009)]{cami2009}
Cami, J., et al. 2009, 
ApJ, 690, L122
\bibitem[Cami et al.(2010a)]{cami2010}
Cami, J., van Malderen, R., \& Markwick, A.J. 2010a, 
ApJS, 187, 409. See http://www.spectrafactory.net/
\bibitem[Cami et al.(2010b)]{cami-C60-2010}
Cami, J., Bernard-Salas, J., Peeters, E., \& Malek, S.E. 2010b, 
Science, 329, 1180
\bibitem[Caris et al.(2004)]{KCl_rot_2004} 
Caris, M., Lewen, F., M{\"u}ller, H.S.P., \& Winnewisser, G.
2004, J. Mol. Struct., 695$-$696, 243
\bibitem[Caris et al.(2009)]{C3H_rot_2009} 
Caris, M., Giesen, T.F., Duan, C., M{\"u}ller, H.S.P., 
Schlemmer, S., \& Yamada, K.M.T. 
2009, J. Mol. Spectrosc., 253, 99
\bibitem[Carvajal et al.(2009)]{13C2_MeFo_rot_2009} 
Carvajal, M., et al. 
2009, A\&A, 500, 1109
\bibitem[Carvajal et al.(2010)]{13C1_MeFo_rot_2010} 
Carvajal, M., Kleiner, I., \& Demaison, J. 
2010, ApJS, 190, 315
\bibitem[Cazzoli \& Puzzarini(2005)]{H13CN_LD_2005} 
Cazzoli, G., \& Puzzarini, C. 
2005, J. Mol. Spectrosc., 233, 280
\bibitem[Cazzoli et al.(2006a)]{DF_rot_2006} 
Cazzoli, G., Puzzarini, C., Tamassia, F., Borri, S., \& Bartalini, S. 
2006a, J. Mol. Spectrosc., 235, 265
\bibitem[Cazzoli et al.(2006b)]{PN_rot_2006} 
Cazzoli, G., Cludi, L., \& Puzzarini, C. 
2006b, J. Mol. Struct., 780, 260
\bibitem[Cazzoli \& Puzzarini(2006)]{PH3_LD_2006} 
Cazzoli, G., \& Puzzarini, C. 
2006, J. Mol. Spectrosc., 239, 64
\bibitem[Cazzoli \& Puzzarini(2008)]{MeC2_4H_LD_2008} 
Cazzoli, G., \& Puzzarini, C. 
2008, A\&A, 487, 1197
\bibitem[Cazzoli et al.(2010a)]{D2O_LD_2010} 
Cazzoli, G., Dore, L., Puzzarini, C., \& Gauss, J. 
2010a, Mol. Phys., 108, 2335
\bibitem[Cazzoli et al.(2010b)]{CF+_rot_2010} 
Cazzoli, G., Cludi, L., Puzzarini, C., \& Gauss, J. 
2010b, A\&A, 509, A1
\bibitem[Chakrabarti \& Tennyson(2006)]{cha06}
Chakrabarti, K., \& Tennyson, J. 2006, 
J. Phys. B., 39, 1485
\bibitem[Chen et al.(2010)]{che10}
Chen, C.-W., Merer, A.J., Chao, J.-M., \& Hsu, Y.-C. 
2010, J. Mol. Spectrosc., 263, 56
\bibitem[Chen et al.(2011)]{che11}
Chen, K.-S., Zhang, G., Merer, A.J., Hsu, J.-C., \& 
Chen, W.-J. 2011, 
J. Mol. Spectrosc., 267, 169
\bibitem[Cheng et al.(2004)]{che04}
Cheng, B.-M., Chung, C.-Y., Bahou. M., Lee, Y.-P., 
Lee, L.C., van Harrevelt, R., \& van Hemert, M.C. 2004, 
J. Chem. Phys., 120, 224
\bibitem[Cheng et al.(2006)]{che06}
Cheng, B.-M., et al. 2006, 
ApJ, 647, 1535
\bibitem[Chowdhury et al.(2006)]{cho06}
Chowdhury, P.K., Merer, A.J., Rixon, S.J., Bernath, P.F., 
\& Ram, R.S. 2006,
Phys. Chem. Chem. Phys., 8, 822
\bibitem[Christen \& M{\"u}ller(2003)]{aGg'_2003}
Christen, D., \& M{\"u}ller, H.S.P.
2003, Phys. Chem. Chem. Phys., 5, 3600
\bibitem[Colin \& Bernath(2010)]{CH2010}
Colin, R., \& Bernath, P.F. 2010, 
J. Mol. Spectrosc., 263, 120
\bibitem[Cushing et al.(2011)]{Y-dwarfs2011}
Cushing, M.C., et al. 2011, 
ApJ, accepted, arXiv:1108.4678
\bibitem[Danielache et al.(2008)]{dan08}
Danielache, S.O., Eskebjerg, C., Johnson, M.S., Ueno, Y., 
\& Yoshida, N. 2008, 
JGRD, 113, 17314
\bibitem[Demyk et al.(2007)]{13C-EtCN_rot_2007} 
Demyk, K., et al.
2007, A\&A, 466, 255
\bibitem[Dickenson et al.(2010)]{dic10}
Dickenson, G.D., Nortje, A.C., Steenkamp, C.M., Rohwer, 
E.G., \& du Plessis, A. 2010,
ApJ, 714, L268
\bibitem[Dore et al.(2009)]{15N-N2H+_rot_2009} 
Dore, L., Bizzocchi, L., Degli Esposti, C., \& Tinti, F. 
2009, A\&A, 496, 275
\bibitem[Dore et al.(2010)]{H2CNH_rot_2010} 
Dore, L., Bizzocchi, L., Degli Esposti, C., \& Gauss, J. 
2010, J. Mol. Spectrosc., 263, 44
\bibitem[Drouin et al.(2006)]{C3H8_rot_2006} 
Drouin, B.J., Pearson, J.C., Walters, A., \& Lattanzi, V. 
2006, J. Mol. Spectrosc., 240, 227
\bibitem[Drouin et al.(2009)]{CH3D_rot_2009} 
Drouin, B.J., Yu, S., Pearson, J.C., \& M{\"u}ller, H.S.P. 
2009, JQSRT, 110, 2077
\bibitem[du Plessis et al.(2006)]{dup06}
du Plessis, A., Rohwer, E.G., \& Steenkamp, C.M. 2006, 
ApJS, 165, 432
\bibitem[du Plessis et al.(2007)]{dup07}
du Plessis, A., Rohwer, E.G., \& Steenkamp, C.M. 2007, 
J. Mol. Spectrosc., 243, 124
\bibitem[Duan et al.(2003)]{MeOD_rot_2003} 
Duan, Y.-B., Ozier, I., Tsunekawa, S., \& Takagi, K. 
2003, J. Mol. Spectrosc., 218, 95
\bibitem[Dubernet et al.(2010)]{VAMDC_2010} 
Dubernet, M.L., Boudon, V., Culhane, J.L., et al. 
2010, JQSRT, 111, 2151
\bibitem[Dulick et al.(2003)]{dul03}
Dulick, M., Bauschlicher, C.W., Burrows, A., Sharp, C.M., 
Sharp, R.S., Ram, R.S., \& Bernath, P.F. 2003, 
ApJ, 594, 651
\bibitem[Eidelsberg \& Rostas(2003)]{eid03}
Eidelsberg, M., \& Rostas, F. 2003, 
ApJS, 145, 89
\bibitem[Eidelsberg et al.(2004a)]{eid04a}
Eidelsberg, M., Launay, F., Ito, K., Matsui, T., Hinnen, 
P.C., Reinhold, E., Ubachs, W., \& Huber, K.P. 2004a, 
J. Chem. Phys., 121, 292
\bibitem[Eidelsberg et al.(2004b)]{eid04b}
Eidelsberg, M. Lemaire, J.L., Fillion, J.H., Rostas, F., 
Federman, S.R., \& Sheffer, Y. 2004b, 
A\&A, 424, 355
\bibitem[Eidelsberg et al. (2006)]{eid06}
Eidelsberg, M., Sheffer, Y., Federman, S.R., Lemaire, J.L., 
Fillion, J.H., Rostas, F., \& Ruiz, J. 2006, 
ApJ, 647, 1543
\bibitem[Elkeurti et al.(2008)]{15NH2D_15NHD2_rot_2008} 
Elkeurti, M., Coudert, L.H., Orphal, J., Wlodarczak, G., 
Fellows, C.E., \& Toumi, S. 
2008, J. Mol. Spectrosc., 251, 90
\bibitem[Endres et al.(2006)]{NHD2_rot_2006} 
Endres, C.P., M{\"u}ller, H.S.P., Br{\"u}nken, S., 
Paveliev, D.G., Giesen, T.F., Schlemmer, S., \& Lewen, F. 
2006, J. Mol. Struct., 795, 242
\bibitem[Endres et al.(2009)]{DME_rot_2009} 
Endres, C.P., Drouin, B.J., Pearson, J.C., M{\"u}ller, H.S.P., 
Lewen, F., Schlemmer, S., \& Giesen, T.~F. 
2009, A\&A, 504, 635
\bibitem[Fillion et al.(2004)]{fil04}
Fillion, J.-H., Ruiz, J., Yang, X.-F., Castillejo, M., 
Rostas, F., \& Lemaire, J.-L. 2004,
J. Chem. Phys., 120, 6531
\bibitem[Fisher et al.(2007)]{18O_MeOH_analysis_2007} 
Fisher, J., Paciga, G., Xu, L.-H., Zhao, S.B., Moruzzi, G., 
\& Lees, R.M. 
2007, J. Mol. Spectrosc., 245, 7
\bibitem[Flores-Mijangos et al.(2004)]{NH_rot_2004} 
Flores-Mijangos, J., Brown, J.M., Matsushima, F., Odashima, H., 
Takagi, K., Zink, L.R., \& Evenson, K.M. 
2004, J. Mol. Spectrosc., 225, 189
\bibitem[Flory et al.(2004)]{CoCl_rot_2004} 
Flory, M.A., Halfen, D.T., \& Ziurys, L~M. 
2004, J. Chem. Phys., 121, 8385
\bibitem[Flory et al.(2006)]{ZnF_rot_2006} 
Flory, M.A., McLamarrah, S.K., \& Ziurys, L.M. 
2006, J. Chem. Phys., 125, 194304
\bibitem[Flory \& Ziurys(2008)]{VN_VO_rot2008} 
Flory, M.A., \& Ziurys, L.M. 
2008, J. Mol. Spectrosc., 247, 76 
\bibitem[Fortman et al.(2010)]{EtCN-T_rot_2010} 
Fortman, S.M., Medvedev, I.R., Neese, C.F., \& De Lucia, F.C. 
2010, ApJ, 725, 1682
\bibitem[Fortman et al.(2011)]{VyCN-T_rot_2011} 
Fortman, S.M., Medvedev, I.R., Neese, C.F., \& De Lucia, F.C. 
2011, ApJ, 737, 20 
\bibitem[Fujimori et al.(2011)]{H2F+_rot_2011} 
Fujimori, R., Kawaguchi, K., \& Amano, T. 
2011, ApJ, 729, L2
\bibitem[Galu\'{e} et al.(2011)]{galue2011}
Galu\'{e}, H.A., Rice, C.A., Steill, J.D., Oomens, J. 2011, 
J. Chem. Phys., 134, 054310
\bibitem[Gendriesch et al.(2009)]{CO_vib_2009} 
Gendriesch, R., Lewen, F., Klapper, G., Menten, K.M., 
Winnewisser, G., Coxon, J.A., \& M\"uller, H.S.P. 
2009, A\&A, 497, 927
\bibitem[Gilijamse et al.(2007)]{gil07}
Gilijamse, J.J., Hoekstra, S., Meek, S.A., Mets\"{a}l\"{a}, 
M., van de Meerakker, S.Y.T., Meijer, G., \& Groenenboom, 
G.C. 2007, 
J. Chem. Phys., 127, 221102
\bibitem[Golubiatnikov et al.(2006)]{H2O_H2O-18_LD_2006} 
Golubiatnikov, G.Y., Markov, V.N., Guarnieri, A., \& Kn{\"o}chel, R. 
2006, J. Mol. Spectrosc., 240, 251
\bibitem[Gottlieb et al.(2007)]{CN-_rot_2007} 
Gottlieb, C.A., Br{\"u}nken, S., McCarthy, M.C., \& Thaddeus, P. 
2007, J. Chem. Phys., 126, 191101
\bibitem[Gottlieb et al.(2010)]{C6H_vib_2010} 
Gottlieb, C.A., McCarthy, M.C., \& Thaddeus, P. 
2010, ApJS, 189, 261
\bibitem[Groner et al.(2008)]{acetone_vib_2008} 
Groner, P., Medvedev, I.R., De Lucia, F.C., \& Drouin, B.J. 
2008, J. Mol. Spectrosc., 251, 180 
\bibitem[Grozdanov et al.(2004)]{gro04}
Grozdanov, T.P., Bouakline, F., Andric, L., \& McCarroll, R. 
2004, J. Phys. B., 37, 1737
\bibitem[Gupta et al.(2007)]{C4H-_C8H-_rot_2007} 
Gupta, H., Br{\"u}nken, S., Tamassia, F., Gottlieb, C.A., 
McCarthy, M.C., \& Thaddeus, P. 
2007, ApJ, 655, L57
\bibitem[Hakalla et al.(2006)]{hak06}
Hakalla, R., Kepa, R., Szajna, W., \& Zachwieja, M. 
2006, Eur. Phys. J. D, 38, 481
\bibitem[Halfen \& Ziurys(2004)]{AlH_D_rot2004} 
Halfen, D.T., \& Ziurys, L.M. 
2004, ApJ, 607, L63
\bibitem[Halfen \& Ziurys(2008)]{MnH_D_rot_2008} 
Halfen, D.T., \& Ziurys, L.M. 
2008, ApJ, 672, L77
\bibitem[Halfen et al.(2008)]{CD_13CH_rot_2008} 
Halfen, D.T., Ziurys, L.M., Pearson, J.C., \& Drouin, B.J. 
2008, ApJ, 687, 731
\bibitem[Halfen et al.(2009a)]{CCP_rot_2009} 
Halfen, D.T., Sun, M., Clouthier, D.J., \& Ziurys, L.M. 
2009a, J. Chem. Phys., 130, 014305
\bibitem[Halfen et al.(2009b)]{VCl_rot_2009} 
Halfen, D.T., Ziurys, L.M., \& Brown, J.M. 
2009b, J. Chem. Phys., 130, 164301
\bibitem[Hargreaves et al.(2011)]{York2011}
Hargreaves, R.J., Li, G., \& Bernath, P.F. 2011, 
ApJ, 735, 111
\bibitem[Harrison \& Brown(2008)]{har08}
Harrison, J.J., \& Brown, J.M. 2008, 
ApJ, 686, 1426
\bibitem[Harrison et al.(2006)]{CrH_analysis_2006} 
Harrison, J.J., Brown, J.M., Halfen, D.T., \& Ziurys, L.M. 
2006, ApJ, 637, 1143
\bibitem[Harrison et al.(2007)]{CoF_rot_2007} 
Harrison, J.J., Brown, J.M., Flory, M.A., Sheridan, P.M., 
McLamarrah, S.K., \& Ziurys, L.M. 
2007, J. Chem. Phys., 127, 194308
\bibitem[Harrison et al.(2008)]{har08a}
Harrison, J.J., Brown, J.M., Chen., J., Steimle, T.C., \& Sears, T.J. 
2008, ApJ, 679, 854
\bibitem[Hase et al.(2010)]{ACE2010}
Hase, F., Wallace, L., McLeod, S.D., Harrison, J.J., \& Bernath, P.F. 2010, 
JQSRT, 111, 521. See http://www.ace.uwaterloo.ca/solaratlas.html/
\bibitem[Haverd et al.(2005)]{hav05}
Haverd, V.E., Lewis, B.R., Gibson, S.T., \& Stark, G. 
2005, J. Chem. Phys., 123, 214304
\bibitem[Heays et al.(2009)]{hea09}
Heays, A.N., Lewis, B.R., Stark, G., Yoshino, K., Smith, 
P.L., Huber, K.P., \& Ito, K. 2009,
J. Chem. Phys., 131, 194308
\bibitem[Hermans et al. (2009)]{her09}
Hermans, C., Vandaele, A.C., \& Fally, S. 2009, 
JQSRT, 110, 756
\bibitem[Hirata et al.(2008)]{alanine_mmw_2008} 
Hirata, Y., Kubota, S., Watanabe, S., Momose, T., \& Kawaguchi, K. 
2008, J. Mol. Spectrosc., 251, 314
\bibitem[Huang et al.( 2011)]{Ames2011}
Huang, X., Schwenke, D.W., \& Lee, T.J. 2011, 
J. Chem. Phys., 134, 044320 and 044321
\bibitem[Huber et al.(2009)]{hub09}
Huber, K.P., Chan, M.-C., Stark, G., Ito, K., \&
Matsui, T. 2009,
J. Chem. Phys., 131, 084301
\bibitem[H{\"u}bers et al.(2009)]{NH+_LMR_2009} 
H{\"u}bers, H.-W., Evenson, K.M., Hill, C., \& Brown, J.M. 
2009, J. Chem. Phys., 131, 034311
\bibitem[Hudgins et al.(2005)]{ames2}
Hudgins, D.M., Bauschlicher, C.W., Allamandola, L.J. 2005, 
ApJ, 632, 316
\bibitem[Iglesias-Groth et al.(2011)]{solid2011}
Iglesias-Groth, S., Cataldo, F., \& Manchado, A. 2011, 
MNRAS, 413, 213
\bibitem[Ilyushin et al.(2004)]{acetamide_rot_2004} 
Ilyushin, V.V., Alekseev, E.A., Dyubko, S.F., Kleiner, I., 
\& Hougen, J.T. 
2004, J. Mol. Spectrosc., 227, 115
\bibitem[Ilyushin et al.(2005)]{glycine_mmw_2005} 
Ilyushin, V.V., Alekseev, E.A., Dyubko, S.F., Motiyenko, R.A., 
\& Lovas, F.J. 
2005, J. Mol. Spectrosc., 231, 15
\bibitem[Ilyushin \& Lovas(2007)]{CH3NH2_2007} 
Ilyushin, V., \& Lovas, F.J. 
2007, J. Phys. Chem. Ref. Data, 36, 1141
\bibitem[Ilyushin et al.(2008)]{HAc_2008} 
Ilyushin, V., Kleiner, I., \& Lovas, F.J. 
2008, J. Phys. Chem. Ref. Data, 37, 97
\bibitem[Ilyushin et al.(2009)]{MeFo_rot_2009} 
Ilyushin, V., Kryvda, A., \& Alekseev, E. 
2009, J. Mol. Spectrosc., 255, 32
\bibitem[Jacquinet-Husson et al.(2011)]{jac11}
Jacquinet-Husson, N., Crepeau, L., Armante, R., et al.
2011, JQSRT, 112, 2395
\bibitem[Jansen et al.(2011)]{me-mp_2011a} 
Jansen, P., Xu, L.-H., Kleiner, I., Ubachs, W., \& Bethlem, H.L. 
2011, Phys. Rev. Lett., 106, 100801
\bibitem[Joblin \& Tielens (2011)]{joblin11}
Joblin, C. \& Tielens, A.G.G.M. (Eds.) 2011,
EAS Publications Series, Vol. 46
\bibitem[Joester et al.(2007)]{joe07}
Joester, J.A., Nakajima, M., Reilly, N.J., Kokkin, D.L., 
Nauta, K., Kable, S.H., \& Schmidt, T.W. 2007, 
J. Chem. Phys., 127, 214303
\bibitem[Jungen et al.(2003)]{jun03}
Jungen, Ch., Huber, K.P., Jungen, M., \& Stark, G. 2003, 
J. Chem. Phys., 118, 4517
\bibitem[Kania et al.(2011)]{TiO2_rot_2011} 
Kania, P., Hermanns, M., Br{\"u}nken, S., M{\"u}ller, H.S.P., 
\& Giesen, T.F. 
2011, J. Mol. Spectrosc., 268, 173
\bibitem[Kassi \& Campargue(2011)]{HD2011}
Kassi, S., \& Campargue, A. 2011, 
J. Mol. Spectrosc., 267, 36
\bibitem[Kawaguchi et al.(1995)]{CW-Leo_survey_1995} 
Kawaguchi, K., Kasai, Y., Ishikawa, S.-I., \& Kaifu, N. 
1995, PASJ, 47, 853
\bibitem[Kawahara et al.(2008)]{kaw08}
Kawahara, H., Kato, H., Hoshino, M., Tanaka, H., \& 
Brunger, M.J. 2008, 
Phys. Rev., A77, 012713
\bibitem[Killian et al.(2007)]{C2H_vib_2007} 
Killian, T.C., Gottlieb, C.A., \& Thaddeus, P. 
2007, J. Chem. Phys., 127, 114320
\bibitem[Kisiel et al.(2009)]{VyCN_rot_2009} 
Kisiel, Z., Pszcz{\'o}{\l}kowski, L., Drouin, B.J., Brauer, 
C.S., Yu, S., \& Pearson, J.C. 
2009, J. Mol. Spectrosc., 258, 26
\bibitem[Kisiel et al.(2010)]{n-PrOH_rot_2010} 
Kisiel, Z., et al. 
2010, Phys. Chem. Chem. Phys., 12, 8329
\bibitem[Kokkin et al.(2007)]{kok07}
Kokkin, D.L., Bacskay, G.B., \& Schmidt, T.W. 2007, 
J. Chem. Phys., 126, 084302
\bibitem[Kryvda et al.(2009)]{formamide_2009} 
Kryvda, A.V., Gerasimov, V.G., Dyubko, S.F., Alekseev, E.A., 
\& Motiyenko, R.A. 
2009, J. Mol. Spectrosc., 254, 28
\bibitem[Lattanzi et al.(2008)]{HCOOH_isos_2008} 
Lattanzi, V., Walters, A., Drouin, B.J., \& Pearson, J.C. 
2008, ApJS, 176, 536
\bibitem[Lattanzi et al.(2010)]{NCO-_rot_2010} 
Lattanzi, V., Gottlieb, C.A., Thaddeus, P., Thorwirth, S., 
\& McCarthy, M.C. 
2010, ApJ, 720, 1717
\bibitem[Lauvergnat et al.(2009)]{CH2DOH_FIR_2009} 
Lauvergnat, D., Coudert, L.H., Klee, S., \& Smirnov, M. 
2009, J. Mol. Spectrosc., 256, 204
\bibitem[Lav\'{i}n et al.(2009)]{lav09}
Lav\'{i}n, C., Velasco, A.M., Mart\'{i}n, I. 2009, 
ApJ, 692, 1354
\bibitem[Lav\'{i}n \& Velasco(2011)]{lav11}
Lav\'{i}n, C., \& Velasco, A.M. 2011,
ApJ, 739, 16
\bibitem[Lefebvre-Brion \& Lewis(2007)]{lef07}
Lefebvre-Brion, H., \& Lewis, B.R., 2007, 
Mol. Phys., 105, 1625
\bibitem[Lefebvre-Brion et al.(2010)]{lef10}
Lefebvre-Brion, H., Liebermann, H.P., \& V\'{a}zquez, 
G.J. 2010,
J. Chem. Phys., 132, 024311
\bibitem[Levshakov et al.(2011)]{me-mp_2011b} 
Levshakov, S.A., Kozlov, M.G., \& Reimers, D. 
2011, ApJ, 738, 26
\bibitem[Lewis et al.(2005a)]{lew05a}
Lewis, B.R., Gibson, S.T., Zhang, W., Lefebvre-Brion, 
H., \& Robbe, J.-M. 2005a, 
J. Chem. Phys., 122, 144302
\bibitem[Lewis et al.(2005b)]{lew05b}
Lewis, B.R., Gibson, S.T., Sprengers, J.P., Ubachs, W., 
Johansson, A., \& Wahlstr\"{o}m, C.-G. 2005b, 
J. Chem. Phys., 123, 236101
\bibitem[Lewis et al.(2008a)]{lew08a}
Lewis, B.R., Baldwin, K.G.H., Sprengers, J.P., Ubachs, 
W., Stark, G., \& Yoshino, K. 2008a, 
J. Chem. Phys., 129, 164305
\bibitem[Lewis et al.(2008b)]{lew08b}
Lewis, B.R., Heays, A.N., Gibson, S.T., Lefebvre-Brion, 
H., \& Lefebvre, R. 2008b, 
J. Chem. Phys., 129, 164306
\bibitem[Li et al.(2011)]{gang2011}
Li, G., Gordon, I.E., Bernath, P.F., \& Rothman, L.S. 2011, 
JQSRT, 112, 1543
\bibitem[Li et al.(2006)]{li06}
Li, W.-B., Zhu, L.-F., Yuan, Z.-S., Liu, X.-J., \& 
Xu, K.-Z. 2006, 
J. Chem. Phys., 125, 154310
\bibitem[Livingston \& Wallace(2003)]{Solar2003}
Livingston, W., \& Wallace, L. 2003, 
\emph{An Atlas of the Solar Spectrum in the Infrared from} 
1850 -- 9000 cm$^{-1}$ (1.1 to 5.4 $\mu$m), \emph{Revised}, 
N.S.O. Technical Report \# 03-001, National Solar Observatory, Tucson. 
See ftp://nsokp.nso.edu/pub/atlas/photatl/
\bibitem[Lovas(2004)]{NIST-RRF_2004}
Lovas, F.J. 
2004, J. Phys. Chem. Ref. Data, 33, 177
\bibitem[Lovas \& Groner(2006)]{13C-acetone_2006} 
Lovas, F.J., \& Groner, P. 
2006, J. Mol. Spectrosc., 236, 173
\bibitem[Lovas et al.(2009)]{1-2-Pr2OH_2009} 
Lovas, F.J., Plusquellic, D.F., Pate, B.H., Neill, J.L., 
Muckle, M.T., \& Remijan, A.J. 
2009, J. Mol. Spectrosc., 257, 82
\bibitem[Maeda et al.(2001)]{TiCl_rot_2001} 
Maeda, A., Hirao, T., Bernath, P.F., \& Amano, T. 
2001, J. Mol. Spectrosc., 210, 250
\bibitem[Maeda et al.(2008)]{H2CS_rot_2008} 
Maeda, A., et al. 
2008, ApJS, 176, 543
\bibitem[Margul{\`e}s et al.(2003)]{HCS+_rot_2003} 
Margul{\`e}s, L., Lewen, F., Winnewisser, G., Botschwina, P., 
\& M{\"u}ller, H.S.P. 
2003, Phys. Chem. Chem. Phys., 5, 2770
\bibitem[Margul{\`e}s et al.(2009a)]{D-Me-MeFo_rot_2009} 
Margul{\`e}s, L., Coudert, L.H., M{\o}llendal, H., Guillemin, 
J.-C., Huet, T.R., \& Jane{\v c}kov{\`a}, R. 
2009a, J. Mol. Spectrosc., 254, 55 
\bibitem[Margul{\`e}s et al.(2009b)]{D_N-EtCN_rot_2009} 
Margul{\`e}s, L., et al.
2009b, A\&A, 493, 565
\bibitem[Margul{\`e}s et al.(2010)]{D-Ac-MeFo_rot_2010} 
Margul{\`e}s, L., et al.
2010, ApJ, 714, 1120
\bibitem[McCall et al.(2003)]{mcc03}
McCall, B.J., Casaes, R.N., \'{A}d\'{a}mkovics, M., \& 
Saykally, R.J. 2003, 
Chem. Phys. Lett., 374, 583
\bibitem[McCarthy \& Thaddeus(2005)]{13C-C3H_etc_2005} 
McCarthy, M.C., \& Thaddeus, P. 
2005, J. Chem. Phys., 122, 174308
\bibitem[McCarthy et al.(2006)]{C6H-_rot_2006} 
McCarthy, M.C., Gottlieb, C.A., Gupta, H., \& Thaddeus, P. 
2006, ApJ, 652, L141
\bibitem[McLamarrah et al.(2005)]{CoO_rot_2005} 
McLamarrah, S.K., Sheridan, P.M., \& Ziurys, L.M. 
2005, Chem. Phys. Lett., 414, 301
\bibitem[Medvedev et al.(2009)]{EtFo_rot_2009} 
Medvedev, I.R., De Lucia, F.C., \& Herbst, E. 
2009, ApJS, 181, 433
\bibitem[Mladenovi{\'c} et al.(2009)]{HONC_rot_2009} 
Mladenovi{\'c}, M., Lewerenz, M., McCarthy, M.C., \& Thaddeus, P. 
2009, J. Chem. Phys., 131, 174308
\bibitem[Mota et al.(2005)]{mot05}
Mota, R., et al. 2005, 
Chem. Phys. Lett., 416, 152
\bibitem[Mukhopadhyay et al.(2002)]{CH2DOH_rot_2002}
Mukhopadhyay, I., Perry, D.S., Duan, Y.-B., Pearson, J.C., Albert, S.,
Butler, R.A.H., Herbst, E., \& De Lucia, F.C.
2002, J. Chem. Phys., 116, 3710
\bibitem[M{\"u}ller \& Christen(2004)]{gGg'_rot_2004} 
M{\"u}ller, H.S.P., \& Christen, D. 
2004, J. Mol. Spectrosc., 228, 298
\bibitem[M{\"u}ller et al.(2004)]{MeOH_rot_2004} 
M{\"u}ller, H.S.P., Menten, K.M., \& M{\"a}der, H. 
2004, A\&A, 428, 1019 
\bibitem[M{\"u}ller \& Br{\"u}nken(2005)]{SO2_rot_2005} 
M{\"u}ller, H.S.P., \& Br{\"u}nken, S. 
2005, J. Mol. Spectrosc., 232, 213
\bibitem[M{\"u}ller et al.(2005)]{CDMS_2}
M{\"u}ller, H.S.P., Schl{\"o}der, F., Stutzki, J.,
\& Winnewisser, G.
2005, J. Mol. Struct., 742, 215
\bibitem[M{\"u}ller et al.(2007)]{SiS_rot} 
M{\"u}ller, H.~S.~P., et al. 
2007, Phys. Chem. Chem. Phys., 9, 1579 
\bibitem[M{\"u}ller et al.(2009)]{MeCN_rot_2009} 
M{\"u}ller, H.S.P., Drouin, B.J., \& Pearson, J.~C. 
2009, A\&A, 506, 1487
\bibitem[M{\"u}ller et al.(2010)]{H2DO+_analyse_2010} 
M{\"u}ller, H.S.P., Dong, F., Nesbitt, D.J., Furuya, T., 
\& Saito, S. 
2010, Phys. Chem. Chem. Phys., 12, 8362 
\bibitem[M{\"u}ller(2010)]{CH+_analyse_2010} 
M{\"u}ller, H.S.P. 
2010, A\&A, 514, L6
\bibitem[M{\"u}ller et al.(2011)]{i-PrCN_rot_2011} 
M{\"u}ller, H.S.P., Coutens, A., Walters, A., Grabow, J.-U., 
\& Schlemmer, S. 
2011, J. Mol. Spectrosc., 267, 100
\bibitem[Nakajima et al.(2009)]{nak09}
Nakajima, M., Joester, J.A., Page, N.I., Reilly, N.J., 
Bacskay, G.B., Schmidt, T.W., \& Kable, S.H. 2009, 
J. Chem. Phys., 131, 044301
\bibitem[Nassar \& Bernath(2003)]{Nassar2003}
Nassar, R., \& Bernath, P.F. 2003, 
JQSRT, 82, 279
\bibitem[Nemes et al.(1994)]{Nemes1994}
Nemes, L. et al. 1994, 
Chem. Phys. Lett., 218, 295
\bibitem[Ozeki et al.(2011)]{CHD_rot_2011} 
Ozeki, H., Bailleux, S., \& Wlodarczak, G. 
2011, A\&A, 527, A64 
\bibitem[Pearson \& Drouin(2006)]{CH+_rot_2006} 
Pearson, J.C., \& Drouin, B.J. 
2006, ApJ, 647, L83 
\bibitem[Pearson et al.(2008)]{EtOH_rot_2008} 
Pearson, J.C., Brauer, C.S., \& Drouin, B.J. 
2008, J. Mol. Spectrosc., 251, 394
\bibitem[Pearson et al.(2009)]{MeOH_vt3_2009} 
Pearson, J.C., Brauer, C.S., Drouin, B.J., \& Xu, L.-H. 
2009, Can. J. Phys., 87, 449
\bibitem[Plusquellic et al.(2009)]{1-3-Pr2OH_2009} 
Plusquellic, D.F., Lovas, F.J., Pate, B.H., Neill, J.L., 
Muckle, M.T., \& Remijan, A.J. 
2009, J. Phys. Chem. A, 1131, 12911
\bibitem[Polehampton et al.(2003)]{17OH_LMR_2003} 
Polehampton, E.T., Brown, J.M., Swinyard, B.M., \& Baluteau, J.-P. 
2003, A\&A, 406, L47
\bibitem[Puzzarini et al.(2003)]{13C17O_13C18O_2003} 
Puzzarini, C., Dore, L., \& Cazzoli, G. 
2003, J. Mol. Spectrosc., 217, 19
\bibitem[Puzzarini et al.(2009)]{H2O-17_LD_2009} 
Puzzarini, C., Cazzoli, G., Harding, M.E., V{\'a}zquez, J., \& Gauss, J. 
2009, J. Chem. Phys., 131, 234304
\bibitem[Ram et al.(2006)]{ram06}
Ram, R.S., Davis, S.P., Wallace, L., Engleman, R., Appadoo, 
D.R.T., \& Bernath, P.F. 2006, 
J. Mol. Spectrosc., 237, 225
\bibitem[Ram \& Bernath(2010)]{NH2010}
Ram, R.S., \& Bernath, P.F. 2010, 
J. Mol. Spectrosc., 260, 115
\bibitem[Ram et al.(2010b)]{ram10b}
Ram, R.S., Wallace, L., \& Bernath, P.F. 2010b, 
J. Mol. Spectrosc., 263, 82
\bibitem[Ram et al.(2010c)]{ram10c}
Ram, R.S., Wallace, L., Hinkle, K., \& Bernath, P.F. 2010c, 
ApJS, 188, 500
\bibitem[Ram \& Bernath(2011a)]{ram11a}
Ram, R.S., \& Bernath, P.F. 2011a, 
ApJS, 194, 34
\bibitem[Ram et al.(2011b)]{ram11b}
Ram, R.S., Tereszchuk, K., Gordon, I.E., Walker, K.A., \& 
Bernath, P.F. 2011b,
J. Mol. Spectrosc., 266, 86
\bibitem[Ricks et al.(2009)]{duncan2009}
Ricks, A.M., Douberly, G.E., Duncan, M.A. 2009, 
ApJ, 702, 301
\bibitem[Rothman et al.(2009)]{HITRAN2008}
Rothman, L.S., et al.
2009, JQSRT, 110, 523. See http://www.cfa.harvard.edu/HITRAN/
\bibitem[Rothman et al.(2010)]{HITEMP2010}
Rothman, L.S., et al.
2010, JQSRT, 111, 2139
\bibitem[Rufus et al.(2003)]{ruf03}
Rufus, J., Stark, G., Smith, P.L., Pickering, J.C., \& 
Thorne, A.P. 2003,
JGRE, 108, 5011
\bibitem[Rufus et al.(2009)]{ruf09}
Rufus, J., Stark, G., Thorne, A.P., Pickering, J.C., 
Blackwell-Whitehead, R.J., Blackie, D., \& Smith, P.L. 
2009, JGRE, 114, 06003
\bibitem[Savin et al.(2011)]{sav11}
Savin, D.W., et al. 2011, 
Prog. Phys., in press
\bibitem[Schmidt \& Bacskay(2007)]{sch07}
Schmidt, T.W., \& Bacskay, G.B. 2007, 
J. Chem. Phys., 127, 234310
\bibitem[Sharpe et al.(2004)]{PNNL2004}
Sharpe, S.W., Johnson, T.J., Sams, R.L., Chu, P.M.,
Rhoderick, G.C., \& Johnson, P.A. 2004, 
Appl. Spectrosc., 58, 1452. See http://nwir.pnl.gov/
\bibitem[Shayesteh \& Bernath(2011)]{sha11}
Shayesteh, A., \& Bernath, P.F. 2011,
J. Chem. Phys., 135, 094308
\bibitem[Shayesteh et al.(2003a)]{BeH2003}
Shayesteh, A., Tereszchuk, K., Bernath, P.F., \& Colin, R. 2003a, 
J. Chem. Phys., 118, 1158
\bibitem[Shayesteh et al.(2003b)]{BeH2-2003}
Shayesteh, A., Tereszchuk, K., Bernath, P.F., \& Colin, R. 2003b, 
J. Chem. Phys., 118, 3622
\bibitem[Shayesteh et al.(2003c)]{MgH2-2003}
Shayesteh, A., Appadoo, D.R.T., Gordon, I., \& Bernath, P.F.  2003c, 
J. Chem. Phys., 119, 7785
\bibitem[Shayesteh et al.(2004a)]{MgH2004}
Shayesteh, A., Appadoo, D.R.T., Gordon, I., LeRoy, R.J., \& Bernath, P.F. 
2004a, J. Chem. Phys., 120, 10002
\bibitem[Shayesteh et al.(2004b)]{CaH2004}
Shayesteh, A., Walker, K.A., Gordon, I., Appadoo, D.R.T., \& Bernath, P.F. 
2004b, J. Mol. Struct., 695-696, 23
\bibitem[Shayesteh et al.(2007)]{sha07}
Shayesteh, A., Henderson, R.D.E., Le Roy, R.J., \& 
Bernath, P.F. 2007,
J. Phys. Chem. A, 111, 12495
\bibitem[Sheffer et al.(2003)]{she03}
Sheffer, Y., Federman, S.R., \& Andersson, B.-G. 2003, 
ApJ, 597, L29
\bibitem[Sheffer \& Federman(2007)]{she07}
Sheffer, Y., \& Federman, S.R. 2007,
ApJ, 659, 1352
\bibitem[Sheridan \& Ziurys(2003)]{NiCN_rot_2003} 
Sheridan, P.M., \& Ziurys, L.M. 
2003, J. Chem. Phys., 118, 6370
\bibitem[Sheridan et al.(2003)]{TiF_rot_2003} 
Sheridan, P.M., McLamarrah, S.K., \& Ziurys, L.M. 
2003, J. Chem. Phys., 119, 9496
\bibitem[Sheridan et al.(2004)]{CoCN_rot_2004} 
Sheridan, P.M., Flory, M.A., \& Ziurys, L.M. 
2004, J. Chem. Phys., 121, 8360
\bibitem[Shi et al.(2010)]{shi10}
Shi, D., Liu, H., Zhang, X., Sun, J., Zhu, Z., \& Liu, Y. 
2010, J. Mol. Struct., 956, 10
\bibitem[Sprengers et al.(2003)]{spr03}
Sprengers, J.P., Ubachs, W., Baldwin, K.G.H., Lewis, B.R., 
\& Tchang-Brillet, W-\"{U}L. 2003, 
J. Chem. Phys., 119, 3160
\bibitem[Sprengers et al.(2004)]{spr04}
Sprengers, J.P., Johansson, A., L'Huillier, A., 
Wahlstr\"{o}m, C.-G., Lewis, B.R., \& Ubachs, W. 2004, 
Chem. Phys. Lett., 389, 348
\bibitem[Sprengers et al.(2005a)]{spr05a}
Sprengers, J.P., Ubachs, W., \& Baldwin, K.G.H. 2005a, 
J. Chem. Phys., 122, 144301
\bibitem[Sprengers et al.(2005b)]{spr05b}
Sprengers, J.P., Reinhold, E., Ubachs, W., Baldwin, K.G.H., 
\& Lewis, B.R. 2005b, 
J. Chem. Phys., 123, 144315
\bibitem[Sprengers \& Ubachs(2006)]{spr06}
Sprengers, J.P., \& Ubachs, W. 2009, 
J. Mol. Spectrosc., 235, 176
\bibitem[Stark et al.(2005)]{sta05}
Stark, G., Huber, K.P., Yoshino, K., Smith, P.L., \& 
Ito, K. 2005, 
J. Chem. Phys., 123, 214303
\bibitem[Stark et al.(2007)]{sta07}
Stark, G., Yoshino, K., Smith, P.L., \& Ito, K. 2007, 
JQSRT, 103, 67
\bibitem[Stark et al.(2008)]{sta08}
Stark, G., Lewis, B.R., Heays, A.N., Yoshino, K., Smith, 
P.L., \& Ito, K. 2008, 
J. Chem. Phys., 128, 114302
\bibitem[Steinmann et al.(2003)]{ste03}
Steinmann, C.M., Rohwer, E.G., \& Stafast, H. 2003, 
ApJ, 590, L123; errat. ApJ, 591, L167
\bibitem[Sumiyoshi et al.(2003)]{CCCl_rot_2003} 
Sumiyoshi, Y., Ueno, T., \& Endo, Y. 
2003, J. Chem. Phys., 119, 1426
\bibitem[Tanabashi et al.(2005)]{tan05}
Tanabashi, A., Hirao, T., Amano, T., \& Bernath, P.F. 
2005, ApJ, 624, 1116
\bibitem[Tanabashi et al.(2007)]{tan07}
Tanabashi, A., Hirao, T., Amano, T., \& Bernath, P.F. 
2007, ApJS, 169, 472
\bibitem[Tashkun \& Perevalov(2011)]{cdsd}
Tashkun, S.A., \& Perevalov, V.I. 2011, 
JQSRT, 112, 1403
\bibitem[Tenenbaum et al.(2007)]{ZnCl_rot_2007} 
Tenenbaum, E.D., Flory, M.A., Pulliam, R.L., \& Ziurys, L.M. 
2007, J. Mol. Spectrosc., 244, 153
\bibitem[Thaddeus et al.(2008)]{C3N-_rot_2008} 
Thaddeus, P., Gottlieb, C.A., Gupta, H., Br{\"u}nken, S., 
McCarthy, M.C., Ag{\'u}ndez, M., Gu{\'e}lin, M., \& Cernicharo, J. 
2008, ApJ, 677, 1132
\bibitem[Thorn et al.(2007)]{tho07}
Thorn, P.A., et al. 2007, 
J. Chem. Phys., 126, 064306
\bibitem[Tinti et al.(2007)]{HCO+_rot_2007} 
Tinti, F., Bizzocchi, L., Degli Esposti, C., \& Dore, L. 
2007, ApJ, 669, L113
\bibitem[Toffoli \& Lucchese(2004)]{tof04}
Toffoli, D., \& Lucchese, R.R. 2004, 
J. Chem. Phys., 120, 6010
\bibitem[V\'{a}zquez et al.(2007)]{vaz07}
V\'{a}zquez, G.J., Amero, J.M., Liebermann, H.P., Buenker, 
R.J., Lefebvre-Brion, H. 2007, 
J. Chem. Phys., 126, 164302
\bibitem[V\'{a}zquez et al.(2009)]{vaz09}
V\'{a}zquez, G.J., Amero, J.M., Liebermann, H.P., \&
Lefebvre-Brion, H. 2009, 
J. Phys. Chem. A, 113, 13395
\bibitem[Vieitez et al.(2008)]{vie08}
Vieitez, M.O., Ivanov, T.I., de Lange, C.A., Ubachs, W., 
Heays, A.N., Lewis, B.R., \& Stark, G. 2008, 
J. Chem. Phys., 128, 134313
\bibitem[Wallace et al.(2001)]{Sunspot2001}
Wallace, L., Hinkle, K., \& Livingston, W.C. 2001, 
\emph{Sunspot Umbral Spectra in the Region} 4000 to 8640 cm$^{-1}$ 
(1.16 to 2.50 $\mu$m), 
N.S.O. Technical Report \# 01-001, National Solar Observatory, Tucson. See 
ftp://nsokp.nso.edu/pub/atlas/spot5atl/
\bibitem[Wallace et al.(2002)]{Sunspot2002}
Wallace, L., Hinkle, K., \& Livingston, W.C. 2002, 
\emph{Sunspot Umbral Spectra in the Regions} 1925 to 2226 and 2392 to 
3480 cm$^{-1}$ (2.87 to 4.18 and 4.48 to 5.35 $\mu$m), 
N.S.O. Technical Report \# 02-001, National Solar Observatory, 
Tucson. See ftp://nsokp.nso.edu/pub/atlas/spot6atl/
\bibitem[Walters et al.(2009)]{DEE_2009} 
Walters, A., M{\"u}ller, H.S.P., Lewen, F., \& Schlemmer, S. 
2009, J. Mol. Spectrosc., 257, 24
\bibitem[Watson(2001)]{wat01}
Watson, J.K.G. 2001, 
ApJ, 555, 472
\bibitem[Weselak et al.(2009)]{wes09}
Weselak, T., Galazutdinov, G.A., Musaev, F.A., Beletsky, 
Y., \& Kre\-{l}owski, J. 2009, 
A\&A, 495, 189
\bibitem[Wu et al.(2007)]{wu07}
Wu, Y.-J., Lu, H.-C., Chen, H.-K., Cheng, B.-M., 
Lee, Y.-P., \& Lee, L.C. 2007, 
J. Chem. Phys., 127, 154311
\bibitem[Xu et al.(2008)]{MeOH_rot_2008} 
Xu, L.-H., et al.
2008, J. Mol. Spectrosc., 251, 305
\bibitem[Yang et al.(2008)]{yan08}
Yang, X., Ben, J., Li, L., \& Chen, Y. 2008, 
JQSRT, 109, 468
\bibitem[Yonezu et al.(2009)]{H2D+_THz_2009} 
Yonezu, T., Matsushima, F., Moriwaki, Y., Takagi, K., 
\& Amano, T. 
2009, J. Mol. Spectrosc., 256, 238
\bibitem[Yoshikawa et al.(2009a)]{C3Cl_rot_2009} 
Yoshikawa, T., Sumiyoshi, Y., \& Endo, Y. 
2009a, J. Chem. Phys., 130, 094302
\bibitem[Yoshikawa et al.(2009b)]{C3F_rot_2009} 
Yoshikawa, T., Sumiyoshi, Y., \& Endo, Y. 
2009b, J. Chem. Phys., 130, 164303
\bibitem[Yu et al.(2009)]{H3O+_2009} 
Yu, S., Drouin, B.J., Pearson, J.C., \& Pickett, H.M. 
2009, ApJS, 180, 119
\bibitem[Yu et al.(2010)]{NH3_rot_2010} 
Yu, S., et al.
2010, J. Chem. Phys., 133, 174317
\bibitem[Yurchenko et al.(2011)]{UCL2011}
Yurchenko, S.N., Barber, R.J., \& Tennyson, J. 2011, MNRAS 413, 1828
\bibitem[Zack et al.(2009)]{ZnO_rot_2009} 
Zack, L.N., Pulliam, R.L., \& Ziurys, L.M. 
2009, J. Mol. Spectrosc., 256, 186
\bibitem[Zelinger et al.(2003)]{HCN_vib_2003} 
Zelinger, Z., Amano, T., Ahrens, V., Br{\"u}nken, S., Lewen, F., 
M{\"u}ller, H.S.P., \& Winnewisser, G. 
2003, J. Mol. Spectrosc., 220, 223
\bibitem[Zhang et al.(2005)]{zha05}
Zhang, G., Chen,  K.-S., Merer, A.J., Hsu, Y.-C., 
Chen, W.-J., Shaji, S., \& Liao, Y.-A. 2005, 
J. Chem. Phys., 122, 244308
\bibitem[Zobov et al.(2008)]{steam2008}
Zobov, N.F., et al. 2008, 
MNRAS, 387, 1093

\end{thebibliography}
\end{document}